\documentclass[sigconf]{acmart}
\usepackage{xcolor}
\usepackage{booktabs}
\usepackage{soul}

\AtBeginDocument{%
  }

\begin{document}

\title{Retrieval Challenges in Low-Resource Public Service Information: A Case Study on Food Pantry Access}

\author{Touseef Hasan}
\affiliation{%
  \institution{Wichita State University}
  \city{Wichita}
  \state{KS}
  \country{USA}
}

\author{Laila Cure}
\affiliation{%
  \institution{Wichita State University}
  \city{Wichita}
  \state{KS}
  \country{USA}
}

\author{Souvika Sarkar}
\affiliation{%
  \institution{Wichita State University}
  \city{Wichita}
  \state{KS}
  \country{USA}
}

\renewcommand{\shortauthors}{Hasan et al.}

\begin{abstract}

Public service information systems are often fragmented, inconsistently formatted, and outdated. These characteristics create low-resource retrieval environments that hinder timely access to critical services. We investigate retrieval challenges in such settings through the domain of food pantry access, a socially urgent problem given persistent food insecurity. We develop an AI-powered conversational retrieval system that scrapes and indexes publicly available pantry data and employs a Retrieval-Augmented Generation (RAG) pipeline to support natural language queries via a web interface. We conduct a pilot evaluation study using community-sourced queries to examine system behavior in realistic scenarios. Our analysis reveals key limitations in retrieval robustness, handling underspecified queries, and grounding over inconsistent knowledge bases. This ongoing work exposes fundamental IR challenges in low-resource environments and motivates future research on robust conversational retrieval to improve access to critical public resources.

\end{abstract}



\keywords{Low-resource environments, Information retrieval, Large language models, Food access, Retrieval-augmented generation}


\maketitle


\section{Introduction and Motivation}

Access to essential public resources increasingly depends on digital information systems. However, these systems are often ineffective because information remains scattered, poorly formatted, and outdated \cite{henninger2013value}. Such characteristics create low-resource environments, where data coverage is incomplete, metadata is noisy or weakly structured, and reliable indexing or benchmarking infrastructure is limited. These constraints are especially problematic in domains where timely access to information directly affects well-being. To study information retrieval (IR) challenges in such environments, we focus on food pantry access—a critical point of interaction between emergency support infrastructure and food-insecure individuals. We conduct a case study in the state of Kansas (U.S.), where 1 in 8 residents—including 1 in 5 children—experience food insecurity\footnote{\url{https://hungerfreekansas.org/}}. In this context, food pantries serve as essential emergency support infrastructure. Yet the online directories listing pantry information are often static, inconsistently structured, and infrequently updated, limiting timely access to available food assistance.

\noindent
In such constrained settings, efficiently utilizing existing resources and delivering accurate information is crucial. We frame this as a low-resource IR challenge at the intersection of social need and fragmented public data ecosystems. Rather than manually browsing static lists, users should be able to issue natural language queries and receive relevant, timely recommendations. To explore this, we develop an AI-powered \emph{Pantry Assistant} that helps users locate food pantries in Kansas through conversational interactions. The system employs a Retrieval-Augmented Generation (RAG) architecture within a web interface to integrate pantry data with Large Language Model (LLM)-based responses. By shifting from static directories to conversational retrieval, this work examines how LLM-based IR systems can improve access to critical public resources in low-resource environments. Beyond food security, our approach can be generalized to other public service domains as well. Our case study highlights broader retrieval challenges in fragmented public service domains and motivates future research on more robust, constraint-aware conversational IR systems.




\section{Description of the Proposed Solution}

\begin{table*}[t]
\centering
\caption{Queries categorized by user requirements, along with a systematic breakdown of LLM performance by query type. An example is provided for each type of query. Accuracy per query type has been calculated for the Llama and GPT models respectively. Notes indicate the most frequent cause of the erroneous responses for each type of user query.}
\vspace{-3mm}
\label{tab:breakdown}
\begin{tabular}{l l r | r r | l}
\toprule
\multicolumn{1}{c}{\textbf{Type of User Query}} &
\multicolumn{1}{c}{\textbf{Example}} &
\multicolumn{1}{c}{\textbf{\# Queries}} &
\multicolumn{2}{|c|}{\textbf{Accuracy}} &
\multicolumn{1}{c}{\textbf{Failure Notes}} \\
\cmidrule(lr){4-5}
& & &
\multicolumn{1}{|c}{\textbf{Llama}} &
\multicolumn{1}{c|}{\textbf{GPT}} &
\\
\midrule
Only location 
& Where can I find food pantries in Sedgwick County? 
& 20 & {0.60} & {0.65} & Retrieving counties \\
Location + hours 
& Are there food pantries in Derby open on Wednesdays? 
& 13 & 0.38 & 0.38 & Retrieving counties \\
Location + ID requirement 
& Find food pantries in Maize that do not require any ID. 
& 13 & 0.15 & 0.30 & ID constraint errors \\
Ambiguous location 
& Where can I get free food near me? 
& 1 & 0.00 & 0.00 & Needs clarification \\
Recall-based exact search 
& Where is the pantry I went to in 67214 on 21st Street?
& 4 & 1.00 & 1.00 & \textit{No errors} \\
\midrule
\textbf{Total} &  & \textbf{50} & \textbf{0.46} & \textbf{0.52} & \\
\bottomrule
\vspace{-5mm}
\end{tabular}
\end{table*}

The Pantry Assistant is built as a modular pipeline combining data preparation, semantic retrieval, and response generation (Figure \ref{fig:workflow}):

\begin{figure}    \centerline{\includegraphics[width=1\textwidth, trim=260 230 40 230, clip]{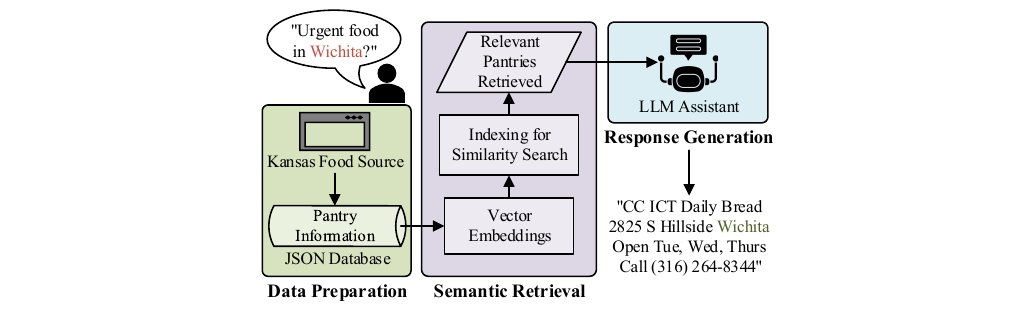}} \vspace{-15px}
    \caption{Our proposed Pantry Assistant. Food pantries are scraped from Kansas Food Source website and stored into a structured JSON database. Data is then embedded and indexed for similarity search to enable semantic retrieval. The retrieved relevant pantries are then passed to an LLM, which recommends pantries to users based on their requirements.} \vspace{-15px}
    \label{fig:workflow}
\end{figure}

\smallskip
\noindent
\textbf{\textit{Data Preparation.}} We identified Kansas Food Source\footnote{\url{https://kansasfoodsource.org/}} (an online directory of over 800 food pantries in Kansas) as the comprehensive source of pantry information. Data from this source was scraped (as of January 31, 2026) and normalized into a structured format. In particular, we parsed the web page of each pantry from the website to extract fields such as name, address, city, county, contact information, hours of operation, eligibility requirements, and other notes, storing this metadata in a standardized JSON schema.

\smallskip
\noindent
\textbf{\textit{Semantic Retrieval.}} We created vector representations of the pantry records to enable efficient semantic retrieval via RAG. We used a sentence-transformer model (all-MiniLM-L6-v2\footnote{\url{https://huggingface.co/sentence-transformers/all-MiniLM-L6-v2}}) to encode metadata of each pantry into an embedding, and indexed these vectors using FAISS \cite{douze2025faiss} for fast similarity search. When a user asks a question, the chatbot identifies key constraints (e.g., location, hours, ID requirements, etc.) and searches the vector index for pantry entries matching those needs. This step is critical to restrict the LLM from recommending hallucinated or irrelevant information to the user. RAG allows the LLM to access and retrieve relevant documents from given data \cite{lewis2020retrieval}, a technique which has been significant in efficient IR in low-resource settings \cite{bikim2025fair, awoleye2025leveraging, telemala2025towards}. The retrieved relevant pantry records are then passed to the LLM.

\smallskip
\noindent
\textbf{\textit{Response Generation.}} The LLM-powered assistant now grounds its response in the pantry data that it retrieves. Our predefined system prompt instructs the LLM to use the data directly and avoid fabrication. The chatbot should also include essential details of the suggested pantries (e.g., open hours, addresses, etc.) to provide a complete recommendation according to the user needs. For generating the response, we tested two LLMs. The first one is an open-source model, Meta Llama 3.1 8B Instruct \cite{dubey2024llama}, chosen for its accessibility and alignment with low-resource deployment. It can be run via a HuggingFace\footnote{\url{https://huggingface.co/}} pipeline free of cost. The second one is GPT-3.5 Turbo \cite{achiam2023gpt}, a proprietary model by OpenAI known for its strong language understanding capabilities. We included both models to compare performance and assess whether an open-source solution can work in a low-resource setting instead of a state-of-the-art proprietary model. The assistant is deployed through a simple web interface built with Gradio\footnote{\url{https://www.gradio.app/}}, making it easy for users to interact with the chatbot in a browser without any installation. This is helpful for users with limited devices or technical skills. We also could quickly share the tool with community partners for feedback.


\section{Preliminary Results and Challenges}

We conducted an initial evaluation of the Pantry Assistant using 50 simulated user queries representing diverse scenarios, including location-specific requests, ambiguous queries, constraint-based questions (e.g., day or ID requirements), and recall-based questions referencing specific addresses or pantry names (Table \ref{tab:breakdown}). A response was considered accurate if it returned correct and relevant information from the curated dataset. Llama 3.1 8B Instruct answered 23/50 queries correctly (46\%), while GPT-3.5 Turbo answered 26/50 (52\%), yielding an average accuracy of 48\%. These results establish an initial baseline, with both models performing comparably.

\noindent
We conducted an error analysis and found that most failures stemmed from three recurring issues: (a) incorrect resolution of county-level location constraints, which led to retrieving pantries outside the intended region; (b) lack of clarification for underspecified queries, particularly those missing location information; and (c) difficulties handling eligibility constraints (e.g., ID requirements) and inconsistencies in knowledge base formatting. These findings highlight retrieval limitations that motivate further improvements.

\noindent


\section{Future Directions and Conclusion}

Our analysis indicates that retrieval robustness, constraint handling, and clarification dialogue are the primary bottlenecks in low-resource conversational IR. Future work will strengthen semantic retrieval through structured constraint modeling (e.g., geographic and eligibility filters), introduce clarification-aware dialogue for underspecified queries, and expand and normalize data sources to improve coverage and consistency. We also plan to conduct more structured user evaluations with community stakeholders. This work frames public service information access as a low-resource IR problem characterized by noisy data, incomplete coverage, and evolving constraints. The Kansas food pantry case study offers a realistic testbed for examining retrieval challenges in socially sensitive domains. Our findings highlight the need for constraint-aware retrieval strategies to improve access to critical public resources.



\bibliographystyle{ACM-Reference-Format}
\bibliography{software}


\end{document}